\documentclass[12pt]{article}
\usepackage{amsfonts,latexsym} 
\begin{document}
\newcommand{\reff}[1]{eq.~(\ref{#1})} 
\newcommand{\Cal}{\mathcal}\newcommand{\de}{\delta}\newcommand{\ga}{\gamma}
\newcommand{\e}{\epsilon} \newcommand{\ot}{\otimes}
\newcommand{\ba}{\begin{array}} \newcommand{\ea}{\end{array}}
\newcommand{\be}{\begin{equation}}\newcommand{\ee}{\end{equation}}
\newcommand{\tmod}{{\cal T}}\newcommand{\amod}{{\cal A}}
\newcommand{\bemod}{{\cal B}}\newcommand{\cmod}{{\cal C}}
\newcommand{\dmod}{{\cal D}}\newcommand{\hmod}{{\cal H}}
\newcommand{\s}{\scriptstyle}\newcommand{\tr}{{\rm tr}}
\newcommand{\einsop}{{\bf 1}}
\newcommand{\bea}{\begin{eqnarray}}
\newcommand{\eea}{\end{eqnarray}}

 
\centerline{\bf \large { EXACT SOLVABILITY IN CONTEMPORARY PHYSICS }}
~~ \\
\centerline{Angela Foerster$^\dagger$, Jon Links$^*$ 
and Huan-Qiang Zhou$^*$.} 
~~\\
\centerline{{$\dagger$}Instituto de F\'{\i}sica da UFRGS, Av. Bento 
Gon\c{c}alves
9500, Porto Alegre, 91501-970, Brasil.}
\centerline {{$*$}Department of Mathematics, The University of 
Queensland, 4072, Australia.}

\begin{abstract}
We review the theory for exactly solving quantum Hamiltonian systems 
through the algebraic Bethe ansatz. We also 
demonstrate how this theory applies to  
current studies in Bose-Einstein condensation and metallic grains
which are of nanoscale size. 
\end{abstract}

\tableofcontents

\vfil\eject

\section{Introduction}

The current realisation of nanotechnology as a viable industry is presenting
a wealth of challenging problems in theoretical physics. Phenomena such as 
Bose-Einstein condensation, entanglement and decoherence in the context of 
quantum information, superconducting correlations
in metallic nanograins, soft condensed matter, 
the quantum Hall effect, nano-optics, the 
Kondo effect and Josephson tunneling phenomena are all emerging to paint a 
vast canvas of interwoven physical theories which provide hope and expectation
that the emergence of new 
nanotechnologies will be rapid in the short term future.
A significant tool in the evolution of the theoretical aspects of these 
studies has been the development and application of potent mathematical 
techniques, which are becoming ever increasingly important as our understanding
of the complexities of these physical systems matures. 

One approach that has recently been raised to prominence in this regard is 
that of the exact solution of a physical model. The necessity of studying the 
exact solution has been demonstrated through the experimental research on
aluminium grains with dimensions at the nanoscale level. The  
work of Ralph, Black and Tinkham (RBT) \cite{rbt} in 1996 
detected the presence of 
superconducting pairing correlations in metallic nanograins which 
manifest as a parity effect in the energy spectrum dependent on whether 
the number of valence electrons on each grain is even or odd. A na\"{\i}ve 
approach to theoretically describe
these systems is to apply the theory of superconductivity 
due to Bardeen, Cooper and Schrieffer (BCS) \cite{bcs}.  
Indeed, the BCS model is the appropriate model for these systems, 
but the associated 
mean field treatment fails. This is because a mean field theory approximates
certain operators in the model by an average value. At the nanoscale
level, the quantum fluctuations are sufficiently large enough that this 
approximation is invalid. In fact, there had been a long harboured
notion that superconductivity would break down for systems where the
mean single particle energy level spacing, which is inversely
proportional to the volume, is comparable to the superconducting gap,
as is in the case of metallic nanograins.
This was conjectured by Anderson \cite{anderson} in 1959 
on the basis of the BCS theory,
but the experiments of RBT show this to not be the case. 
Consequently, an exact solution is highly desired,
a view that has been promoted in \cite{nature}.

The study of exact solutions of quantum mechanical 
models has its origins in the 
work of Bethe in 1931 on the Heisenberg model \cite{bethe}. The field 
received a tremendous impetus
in the 1960s with the work of McGuire \cite{jim}, 
Yang \cite{yang}, Baxter \cite{b} and Lieb and Wu \cite{lw}, 
and has prospered ever since. The work of RBT cited above
has brought the discipline to a new audience,  
when it was realised that the exact
solution of the BCS model had been obtained, though largely ignored, by
Richardson in 1963 \cite{richardson}. 
The reason that Richardson's work was overlooked for so long is because
the theory that had been proposed by BCS was so spectacularly successful
that there had never been a need to use an alternative approach.  
Once the results of RBT were communicated however, it was clear that a
new viewpoint was needed. When the condensed matter physics community
became aware of Richardson's work, his results were promptly adopted and 
it was shown that the analysis of  the exact solution  
gave agreement with the experiments \cite{vdr}. A concise yet
informative account of the
developments is given in \cite{s}. 

In this review we will recount the quantum inverse scattering method 
and the associated algebraic Bethe ansatz method for the exact solution
of integrable quantum Hamiltonians. We then show
how this procedure can be applied for the analysis of three models which 
are the focus of many current theoretical studies; 
a model for two Bose-Einstein 
condensates coupled via Josephson tunneling, a model for atomic-molecular 
Bose-Einstein condensation and the BCS model. In each case we undertake 
an asymptotic analysis of the solution and demonstrate how this can 
be applied to extract the asymptotic behaviour of certain correlation
functions at zero temperature through use of the Hellmann-Feynman
theorem \cite{hf}.



\def\a{\alpha}
\def\b{\beta}
\def\d{\dagger}
\def\e{\epsilon}
\def\g{\gamma}
\def\K{\kappa}
\def\l{\lambda}
\def\o{\omega}
\def\t{\theta}
\def\s{\sigma}
\def\D{\Delta}
\def\L{\Lambda}
\def\ap{\approx} 


\def\beq{\begin{equation}}
\def\eeq{\end{equation}}
\def\bea{\begin{eqnarray}}
\def\eea{\end{eqnarray}}
\def\ba{\begin{array}}
\def\ea{\end{array}}
\def\no{\nonumber}
\def\le{\langle}
\def\re{\rangle}
\def\lt{\left}
\def\rt{\right}
\def\o{\omega}
\def\d{\dagger}
\def\nn{\nonumber} 
\def\j{{ {\cal J}}}
\def\n{{\hat n}}
\def\N{{\hat N}}

\section{Quantum inverse scattering method}

First we will review the basic features of the quantum inverse scattering
method \cite{kib,faddeev}. 
The theory of exactly solvable quantum systems in this 
setting relies on the existence of a solution 
$R(u)\in {\rm End} (V\otimes V)$, where $V$ denotes a vector space,  
which satisfies the Yang-Baxter equation 
acting on the three-fold tensor product space $V\otimes V\otimes V$
\beq
R _{12} (u-v)  R _{13} (u)  R _{23} (v) =
R _{23} (v)  R _{13}(u)  R _{12} (u-v). 
\label{ybe} \eeq
Here $R_{jk}(u)$ denotes the matrix in ${\rm End}
(V \otimes V\otimes V)$ acting non-trivially on the
$j$-th and $k$-th spaces and as the identity on the remaining space.
The $R$-matrix solution may be viewed as the structural constants for the
Yang-Baxter algebra which is generated by the monodromy matrix $T(u)$
whose entries generate the algebra
\beq
R_{12}(u-v) T_1(u) T_2(v)=
T_2(v) T_1(u)R_{12}(u-v). \label{yba}
\eeq
We note that as a result of (\ref{ybe}) the Yang-Baxter is necessarily 
associative.
In component form we may write
$$\sum_{p,q}R^{pq}_{ik}(u-v)T_p^j(u)T_q^l(v)=
\sum_{p,q}T^p_k(v)T^q_i(u)R_{qp}^{jl}(u-v) $$
so the $R^{kl}_{ij}(u)$ give the structure constants of the
algebra.

Here, we will only concern ourselves with the $su(2)$ invariant $R$-matrix
which has the form
\bea
R(u)& = & \frac{1}{u+\eta}({u.I\otimes I+\eta P})\nn \\
&=& \left ( \begin {array} {cccc}
1&0&0&0\\
0&b(u)&c(u)&0\\
0&c(u)&b(u)&0\\
0&0&0&1\\
\end {array} \right ),
\label{rm} \eea
with $b(u)=u/(u+\eta)$ and 
$c(u)=\eta/(u+\eta)$. Above, $P$ is the permutation operator which satisfies 
$$P(x\otimes y)=y\otimes x~~~~\forall\,x,\,y\in\,V.$$ 
In this case the Yang-Baxter algebra has four elements which we express as
\beq
T(u) = \left ( \begin {array} {cc}
A(u)&B(u)\\
C(u)&D(u)
\end {array} \right ). \label{mono}
\eeq

Next suppose that we have a representation, which we denote $\pi$, of the 
Yang-Baxter algebra. For later convenience we set 
$$L(u)=\pi\left( T(u)\right) $$ 
which we refer to as an $L$-operator.   
Defining the transfer matrix through 
\beq 
t(u) = {\rm tr}\, \pi\left((T(u))\right)=\pi\left(A(u)+D(u)\right)
\label{tm} \eeq  
it follows from (\ref{ybe}) that the 
transfer matrices commute for different values of the spectral parameters;
viz. 
\beq [t(u),\,t(v)]=0~~~\forall u,v. \label{ctm} \eeq  
There are two significant consequences of (\ref{ctm}). The first is that 
$t(u)$ may be diagonalised independently of $u$, that is the eigenvectors 
of $t(u)$ do not depend on $u$. Secondly, taking a series expansion
$$t(u)=\sum_kc_ku^k $$
it follows that
$$[c_k,\,c_j]=0 ~~~~\forall\,k,\,j.$$
Thus for any Hamiltonian which is expressible as a function of the 
the operators $c_k$ only, then each $c_k$ corresponds to
an operator representing a constant of the motion since it will commute 
with the Hamiltonian. 
When the number of conserved quantities is equal to the number of
degrees of freedom of the system, the model is said to be integrable.

An important property of the Yang-Baxter algebra is that it has a 
co-multiplication structure which allows us to build tensor product 
representations. In particular, given two $L$-operators $L^U,\,L^W$ 
acting on $V\otimes U$ and $V\otimes W$ respectively, 
then $L=L^UL^W$ is also an $L$-operator as can be see from 
\bea R_{12}(u-v)L_1(u)L_2(v)&=&R_{12}(u-v)L^U_1(u)L^W_1L^U_2(v)L^W_2(v)\nn \\
&=&R_{12}(u-v)L^U_1(u)L^U_2(v)L^W_1(u)L^W_2(v) \nn \\
&=&L^U_2(v)L^U_1(u)R_{12}(u-v)L^W_1(u)L^W_2(v) \nn \\
&=&L^U_2(v)L^U_1(u)L^W_2(v)L^W_1(u)R_{12}(u-v) \nn \\
&=&L^U_2(v)L^W_2(v)L^U_1(u)L^W_1(u)R_{12}(u-v) \nn \\
&=&L_2(v)L_1(u)R_{12}(u-v). \nn \eea 
Furthermore, if $L(u)$ is an $L$-operator then so is $L(u+\alpha)$ for any 
$\alpha$ since the $R$-matrix depends only on the difference of the 
spectral parameters.  

\subsection{Realisations of the Yang-Baxter algebra}

In order to construct a specific model, we must address the question of 
determining a realisation of the Yang-Baxter algebra. Here we will present 
several examples which will all be utilised later. The first realisation
comes from the $R$-matrix itself, since it is apparent from (\ref{ybe}) 
that we 
can make the identification $L(u)=R(u)$ such that a representation of 
(\ref{yba}) is obtained.
This is the realisation used in the construction of the Heisenberg model 
\cite{kib,faddeev}. 
A second realisation is given by $L(u)=G$ ($c$-number realisation), 
where $G$ is an arbitrary $2\times 
2$ matrix whose entries do not depend on $u$. 
This follows from the fact that $[R(u),\,G\otimes G]=0$.
 
There is a realisation 
in terms of canonical boson operators $b,\,b^{\dagger}$ with the relations
$[b,\,b^{\dagger}]=1$ which reads
\cite{kt}
\beq 
L^{b}(u) = \left ( \begin {array} {cc}
u+\eta \N &b\\
b^\dagger & \eta ^{-1}\\
\end {array} \right )
\label{lb} \eeq 
where $\N=b^{\dagger}b$. 
There also exists a realisation in terms of the $su(2)$ Lie algebra 
 with generators $S^z$ and $S^{\pm}$ \cite{kib,faddeev},
\beq 
 L^S(u) = \frac{1}{u}\left ( \begin {array} {cc}
 u-\eta S^z &-\eta S^+\\
 -\eta S^-& u+\eta S^z\\
 \end {array} \right ),
\label{ls} \eeq  
 with the commutation relations $[S^z, S^{\pm}] = \pm S^\pm, [S^+,S^-]
 =2 S^z$. It is worth noting that in the case when the $su(2)$ algebra
takes the spin 1/2 representation the resulting $L$-operator is equivalent
to that given by the $R$-matrix.  
 Another is realised in terms of the  $su(1,1)$
 generators $K^z$ and $K^{\pm}$ \cite{jurco,rybin},
\beq 
 L^K(u) = \left ( \begin {array} {cc}
 u+\eta K^z &\eta K^-\\
 -\eta K^+& u-\eta K^z\\
 \end {array} \right ),
\label{lk} \eeq  
 with the commutation relations $[K^z, K^{\pm}] = \pm K^\pm, [K^+,K^-]
 =-2 K^z$. 

Below we will use these realisations to construct a variety of exactly 
solvable models. First however, we will introduce the algebraic Bethe 
ansatz which provides the exact solution.

\section{Algebraic Bethe ansatz method of solution}

For a given realisation of the Yang-Baxter algebra, 
the solution to the problem of finding the eigenvalues of the 
transfer matrix (\ref{tm}) via the algebraic Bethe ansatz is obtained by 
utilising the commutation relations of the Yang-Baxter algebra. 
We have from the defining relations (\ref{yba}) that (among other
relations)
\bea
& & [A(u), A(v)] =
    [D(u), D(v)] = 0, \no\\
& & [B(u), B(v)] =
    [C(u), C(v)] = 0, \no\\
& & A(u)C(v) =
\frac {u-v+\eta}{u-v} C(v) A(u)-\frac {\eta}{u-v}
C(u) A(v), \no\\
& & D (u) C(v) =
\frac {u-v-\eta}{u-v} C(v) D(u) +\frac {\eta}{u-v}
C(u) D(v). \label{rels}   
\eea

A key step in successfully applying the algebraic Bethe ansatz approach
is finding
a suitable pseudovacuum state, $\left|0\right>$, which has the properties
\bea A(u)\left|0\right>&=&a(u)\left|0\right>, \nn \\
B(u)\left|0\right>&=&0, \nn \\ 
C(u)\left|0\right>&\neq & 0, \nn \\
D(u)\left|0\right>&=&d(u)\left|0\right> \nn \eea 
where $a(u)$ and $d(u)$ are scalar functions. 

Assuming the existence of such a pseudovacuum state, choose the Bethe state 
\beq \left|\vec v\right>\equiv\left|v_1,...,v_M\right>
= \prod ^M_{i =1} C(v_i) \left|0 \right>. \label{state} \eeq 
Note that because $[C(u),\,C(v)]=0$, the ordering is not important in
(\ref{state}). 
The approach of the algebraic Bethe ansatz is to
use the relations (\ref{rels}) to determine the action of $t(u)$ on
$\left|\vec v\right>$. The result is 
\bea  
t(u) \left|\vec v\right> 
&=& \L (u,\,\vec v) 
\left|\vec v\right>  \nn \\
&&~~-\left(\sum_i^N\frac{\eta a(v_i)}{u-v_i}\prod_{j\neq i}^{M}
\frac{v_i-v_j+\eta}{v_i-v_j}
\right)\left|v_1,...v_{i-1},u,v_{i+1},...,v_M\right> \nn \\ 
&&~~+\left(\sum_\a^M\frac{\eta d(v_i)}{u-v_i}\prod_{j\neq i}^{M}
\frac{v_i-v_j-\eta}{v_i-v_j}
\right)\left|v_1,...v_{i-1},u,v_{i+1},...,v_M\right> \label{osba}  
\eea 
where
\beq 
\L(u,\,\vec v) = a(u) \prod ^M_{i=1} 
\frac {u-v_i+\eta}
{u-v_i}+
d(u) \prod ^M_{i=1} \frac {u-v_i-\eta}
{u-v_i}.  
\label{tme} \eeq 
The above shows that $\left|\vec v\right>$ becomes an eigenstate of the
transfer matrix with eigenvalue (\ref{tme}) whenever
the Bethe ansatz equations 
\beq
\frac{a(v_i)}{d(v_i)}=
\prod ^M_{j \neq i}\frac {v_i -v_j - \eta}{v_i -v_j +\eta},
~~~~~~~i=1,...,M.\label{bae} \eeq
are satisfied. 
Note that in the derivation of the Bethe ansatz equations
 it is required that $v_i\neq v_j\,
\forall\,i,\,j.$ 
This is a result of the Pauli Principle for Bethe 
ansatz solvable models as developed in \cite{ik82} for the
Bose gas. 
We will not reproduce the proofs for the present cases, as they follow
essentially the same argument as \cite{ik82}.  

\subsection{Scalar products of states}

One of the important applications of the above discussion 
is that there
exists a  formula due to Slavnov \cite{kib,slavnov,kmt} 
for the scalar product of states 
obtained via the algebraic Bethe ansatz for the $R$-matrix (\ref{rm}). 
The formula reads 
\bea S(\vec v:\,\vec u)&=&\left<0\right|B(v_1)...B(v_M)C(u_1)...C(u_M)
\left|0\right> \nn \\
&=&\frac{{\rm det} F(\vec u:\,\vec v)}
{{\rm det} V(\vec u:\, \vec v)}
\nn \eea 
where 
$$F_{ij}=\frac{\partial}{\partial v_i} \Lambda( u_j,\,{\vec v}), ~~~~
V_{ij}=\frac{1}{u_j-v_i}, $$ 
the parameters $\{v_i\}$ satisfy the Bethe ansatz equations (\ref{bae}),
and $\{u_j\}$ are arbitrary. The significance of this result is that 
it opens the possibility to determine form factors and correlation functions
for any model which can be derived in this manner. Although we will not 
go into any details here, we wish to point out that explicit results 
for two of the the models which we will discuss subsequently can be found in 
\cite{zlmg,lz}.

\section{A model for two coupled Bose-Einstein condensates}



\def\a{\alpha}
\def\b{\beta}
\def\d{\dagger}
\def\e{\epsilon}
\def\g{\gamma}
\def\K{\kappa}
\def\ap{\approx}
\def\l{\lambda}
\def\o{\omega}
\def\t{\tilde{\tau}}
\def\s{S}
\def\D{\Delta}
\def\L{\Lambda}
\def\T{{\cal T}}
\def\TT{{\tilde{\cal T}}}

\def\beq{\begin{equation}}
\def\eeq{\end{equation}}
\def\bea{\begin{eqnarray}}
\def\eea{\end{eqnarray}}
\def\ba{\begin{array}}
\def\ea{\end{array}}
\def\no{\nonumber}
\def\le{\langle}
\def\re{\rangle}
\def\lt{\left}
\def\rt{\right}

Experimental realisation of  Bose-Einstein condensates in
dilute atomic alkali gases has stimulated 
a diverse range of theoretical and experimental  
research activity \cite{pw98,dgps99,l01,otfyk,cbfmmtsi}. 
A particularly exciting possibility is that
a pair of Bose-Einstein condensates (such as a Bose-Einstein condensate
trapped in a
double-well potential) may provide a model tunable system
in which to observe macroscopic quantum tunneling.
Below we will show that a model 
Hamiltonian for a pair of
coupled Bose-Einstein condensates admits an exact solution.
The model is also realisable in  Josephson coupled 
superconducting metallic nanoparticles \cite{schon}, which has applications
in the implementation of solid state quantum computers. 

The canonical Hamiltonian which describes tunneling between 
two Bose-Einstein condensates takes the form \cite{l01}  
\bea
H&=& \frac {K}{8}  (N_1- N_2)^2 - \frac {\Delta \mu}{2} (N_1 -N_2)
 -\frac {\cal {E} _J}{2} (b_1^\dagger b_2 + b_2^\dagger b_1).
\label {ham} \eea
where $b_1^\dagger, b_2^\dagger$ denote the single-particle creation
operators in the two wells and  $N_1 = b_1^\dagger b_1, 
N_2 = b_2^\dagger b_2$ are the corresponding
boson number operators. The total boson number $N_1+N_2$
is conserved and set to the fixed value of $N$. 
The physical meaning of the coupling parameters
for different realisable systems 
may be found in  
\cite{l01}. 
It is useful to divide the parameter
space into three regimes; viz. Rabi ($K/{\cal E}_J<< N^{-1}$),
Josephson ($N^{-1}<<K/{\cal E}_J<<N$) and Fock ($N<<K/{\cal E}_J $).   
There is a correspondence between
(\ref{ham}) and the motion of a pendulum \cite{l01}. In the Rabi and Josephson
regimes this motion is semiclassical, unlike the case of the Fock regime.
For both the Fock and Josephson regimes the analogy corresponds to a 
pendulum with fixed length, while in the Rabi regime the length varies.  
An important problem is to study the behaviour in the crossover regimes, which 
is accessible through the exact solution.
The exact solvability of (\ref{ham}) which we discuss here follows from the
fact that it is mathematically equivalent to the discrete self-trapping
dimer model  
studied by Enol'skii
et al. \cite{esks91}, who solved the model through the algebraic Bethe
ansatz. Below we will describe this construction.

The co-multiplication behind the Yang-Baxter
algebra allows us to choose the following 
representation of the monodromy matrix
\bea
L(u)& =&  L^{b}_1(u+\omega) L^{b}_2(u-\omega) \no \\
&=& \pmatrix{ (u+\o+\eta N_1)(u-\o+\eta N_2)+b_2^{\d}b_1 &
(u+\o+\eta N_1)b_2+\eta^{-1}b_1 \cr 
(u-\o+\eta N_2)b_1^{\d}+\eta^{-1}b^{\d}_2 & b_1^{\d}b_2+\eta^{-2} }
.\label{real} \eea 
Defining the transfer matrix as before through 
$t(u) = {\rm tr}\, (L(u))$ 
we have explicitly in the present case 
\bea 
t(u) &=& u^2 + u \eta \N 
+ \eta^2 N_1N_2 + \eta \omega(N_2-N_1)
+b_2^\dagger b_1+ b_1^\dagger b_2 + \eta^{-2} - \omega^2.
\nn \eea 
Then 
$$ t^{\prime}(0)= \left.\frac{dt}{du}\right|_{u=0}= 
\eta\N $$
and it is easy to verify that the Hamiltonian is related with the transfer
matrix $t(u)$ by 
$$
H=-\K \left (t(u) -\frac {1}{4} (t'(0))^2-
u t'(0)-\eta^{-2}
+\omega^2 -u^2\right),  
$$
where the following identification has been made for the coupling constants 
\bea
\frac {K}{4} &=&  \frac {\K \eta^2}{2}, ~~~ 
\frac {\Delta \mu}{2} =  -\K \eta \omega, ~~~
\frac {\cal {E}_J}{2} =  \K . \no
\eea

An explicit representation of (\ref{mono}) is obtained from (\ref{real})
with the identification
\bea 
A(u)&=& (u+\o+\eta N_1)(u-\o+\eta N_2)+b_2^{\d}b_1 \nn \\
B(u)&=&
(u+\o+\eta N_1)b_2+\eta^{-1}b_1 \nn \\
C(u)&=&
(u-\o+\eta N_2)b_1^{\d}+\eta^{-1}b^{\d}_2 \nn \\
D(u)&=& b_1^{\d}b_2+\eta^{-2}. \nn \eea  
Choosing the 
Fock vacuum as the pseudovacuum, 
which satisfies $B(u)\left|0\right>=0$ as required 
by the Bethe ansatz procedure, the eigenvalues $a(u)$ and $d(u)$
of $A(u)$ and $D(u)$ on $\left|0\right>$ are 
\bea
a(u)  &=& (u+\omega)(u-\omega),\no\\
d(u)  &=& \eta^{-2}.\nn
\eea
The Bethe ansatz equations are then explicitly 
\beq
\eta^2 (v^2_i -\omega^2)=
\prod ^N_{j \neq i}\frac {v_i -v_j - \eta}{v_i -v_j +\eta}
\label{becbae} \eeq
with the eigenstates of the form (\ref{state}) with $C(u)$ given as above. 
{}From the Bethe ansatz equations, we may derive the useful identity
\bea
\prod_{i=1}^m\eta^2(v^2_i-\o^2)=\prod_{i=1}^m\prod_{j=m+1}^N
\frac{v_i-v_j-\eta}{v_i-v_j+\eta} \label{id} \eea
which will be used later.

It is clear that the Bethe states are eigenstates of $\hat N$
with eigenvalue $N$.
As $N$ is the total number of bosons,
we expect $N+1$ solutions of the Bethe ansatz equations. 
As mentioned earlier, we must exclude any solution in which the 
roots of the Bethe ansatz equations are not distinct. 
For example, the solution
\beq v_j=\pm\sqrt{\o^2-(-1)^N\eta^{-2}},~~~~\forall j \label{nonsol}\eeq 
of (\ref{becbae}) is invalid, except when $N=1$. (Note the error in
\cite{lz}). 
For a given valid solution of 
the Bethe ansatz equations, the energy of the
Hamiltonian is obtained from the transfer matrix eigenvalues 
(\ref{tme}) and reads 
\bea E
&=&-\K\left(\eta^{-2}\prod_{i=1}^N(1+\frac{\eta}{v_i-u})
-\frac{\eta^2N^2}{4} -u\eta N-u^2 
\right. \no \\
&&~~~~~~\left.
-\eta^{-2}+\o^2+(u^2-\o^2)\prod_{i=1}^N(1-\frac{\eta}{v_i-u})
\right).     \label{becnrg} \eea   
Note that this expression is independent of the spectral parameter $u$ which 
can be chosen arbitrarily. The formula simplifies considerably with the 
choice $u=\omega$, by employing (\ref{id}), which yields a polynomial 
form.   
\bea
E&=&-\K\left(\eta^{-2}\prod_{i=1}^N\eta^2(v_{i}-\omega+\eta)
(v_{i}+\omega) 
-\frac{\eta^2N^2}{4}
-\eta\o N-\eta^{-2}\right). \no    
\eea 
However, for the purpose of an asymptotic analysis 
in the Rabi regime, 
it is more convenient to choose 
$u=0$, 
while for the Fock regime we use $u=\eta^2$.

\subsection{Asymptotic analysis of the solution} 

Here we will recall the asymptotic analysis of the exact solution that 
was conducted in  \cite{zlmx}. 
We start the analysis with the Rabi regime where 
$\eta ^2 N <<1$.   
{}From the Bethe ansatz equations
 it is clear that $\eta^2v_{i}^2\rightarrow 1$ as 
$\eta\rightarrow 0$, so that 
$v_{i}\approx\pm \eta^{-1}$. 
However, 
when $\eta=0$ we know that the Hamiltonian is diagonalisable 
by using the Bogoliubov transformation, 
from which we can deduce 
that the solution of the Bethe ansatz equations
 corresponding to the ground state 
must have $v_{i}\approx\eta^{-1}$. 
Therefore it is reasonable to
consider the asymptotic expansion
\beq 
v_{i}\approx\eta^{-1}+\e_{i}+\eta\delta_i. \label{exp} 
\eeq 
Excitations correspond to changing the signs of the 
leading terms in the Bethe ansatz roots. To study the asymptotic
behaviour for the $m$th excited state, we set 
\bea v_i&\approx&-\eta^{-1}+\e_i+\eta\delta_i,~~~~~~i=1,...,m, \no \\ 
v_i&\approx& \eta^{-1}+\e_i+\eta\delta_i,   ~~~~~~~i=m+1,...,N, 
\label{exp1} \eea 
with the convention that the ground state corresponds to $m=0$.

{}From the leading terms of the Bethe ansatz equations
 for $v_i,\,i\leq m$ we find 
\beq
\e_i=\sum_{j\neq i}^{m}\frac{1}{\e_i-\e_j}, \label{eps1}
\eeq
which implies
$$\sum_{i=1}^{m}\e_i=0,~~~~\sum_{i=1}^{m}\e_i^2=\frac{m(m-1)}{2}.
$$
In a similar fashion we have for $m<i\leq N$ 
\beq
\e_i=-\sum_{\stackrel{j=m+1}{j\neq i}}^{N}\frac{1}{\e_i-\e_j}, \label{eps2}
\eeq
which implies
$$\sum_{i=m+1}^{N}\e_i=0,~~~~\sum_{i=m+1}^{N}\e_i^2
=-\frac{(N-m)(N-m-1)}{2}.  $$

It is clear from (\ref{eps1}) and (\ref{eps2})
why the Pauli exclusion principle applies in the 
present case. In the asymptotic expansion for $v_i$, $\e_i$ is assumed
finite.  
However, if $v_i=v_j$ for some $i,\,j$, then $\e_i=\e_j$ and 
(\ref{eps1}) and (\ref{eps2})
imply that $\e_i,\,\e_j$ are infinite which is a contradiction. Hence 
$v_i$ must be distinct for different $i$. Note also that  
for this approximation to
be valid we require $\eta^{-1}>> \e_i.$ However, we
see that $|\e_i|$ is of the order of $N^{1/2}$. Thus our approximation
will be valid for $\eta N^{1/2}<<1$, which is precisely the criterion 
for the Rabi region, 
and consequently
$N$ cannot be arbitrarily
large for fixed $\eta$, or vice versa. 

Now we go to the next order. 
{}From (\ref{id})  we find 
\bea
\sum_{i=1}^m\delta_i 
&=&-\frac{m(m-1)}{4}+\frac{m(m-N)}{2}
-\frac{m\o^2}{2} 
\no \eea
\bea 
\sum_{i=m+1}^{N}\delta_i 
&=&-\frac{(N-m)(N-m-1)}{4}+\frac{m(m-N)}{2}
+\frac{(N-m)\o^2}{2}  \no \eea   
which using  (\ref{becnrg}) leads us to the result 
$$
\frac{E_m}{\kappa} \ap -N+2m-\frac{\eta^2\o^2(N-2m)}{2}
+\frac{\eta^2N}{4}+\frac{\eta^2}{2}m(N-m).
$$
The energy level spacings $\Delta_m=E_m-E_{m-1}$ are thus 
\bea 
\Delta_m 
&\ap&\kappa\left(2+\eta^2\o^2+\frac{\eta^2}{2}(N-2m+1)\right). \no \eea 
One may check that $\Delta_m /N$ is of the order of $N^{-1}$. 
This indicates that the Rabi regime is semiclassical \cite{l01}.
This value for the gap between the ground and first excited state 
agrees,
to leading order in $\eta^2 N$,  
 with the Gross-Pitaevskii mean-field theory \cite{sfgs97}
giving a Josephson plasma frequency of
$\omega_J = 2 \kappa ( 1 + \eta^2 N/2)^{1/2}$.

Now we look at the asymptotic behaviour of the Bethe ansatz equations
 in the Fock regime
$\eta^2 >> N$.  It is necessary to 
distinguish the following cases: (i) $\o=0$ and (ii) $\o \neq
0$.

(i) $\o = 0$. In this case, it is appropriate to consider the permutation
operator $P$ which interchanges the labels 1 and 2 in (\ref{ham}). 
For $\o=0$, $P$ commutes with the Hamiltonian, and any eigenvector of
the Hamiltonian is also an eigenvector of $P$ with
eigenvalue $\pm 1$. Therefore the Hilbert space splits into the direct
sum of two subspaces corresponding to the symmetric and antisymmetric
wavefunctions. From now on we restrict ourselves to the case when $N$ is
even, i.e., $N=2M$, although a similar calculation is also applicable to 
the case when $N$ 
is odd. A careful analysis leads us to conclude that the ground
state lies in the symmetric subspace. The asymptotic form of
the roots of the Bethe ansatz equations  
for the ground state takes the ``string''-like structure
$$
v_{j\pm} \ap -(M-j) \eta \pm i \frac {C^j_M}{(j-1)!} \eta ^{-(2j-1)} 
+
M(M+1) \eta ^{-3} \delta _{j1}, 
~~~~~~~~j=1,\cdots, M $$  
where $C^j_M$ is a binomial coefficient.  For
this asymptotic ansatz to be valid, we require that any term in the
asymptotic expansion should be much smaller than those preceeding. This 
yields $\eta^2 >> N$ which coincides with  the defining condition for the Fock
region. Throughout, the Pauli exclusion principle
has been taken into account to exclude any possible spurious solutions 
of the Bethe ansatz equations.  

The above structure  clearly indicates that in the ground state 
the $N$ bosons fuse into
$M$ ``bound'' states and excitations correspond to a breakdown of these 
bound states. Specifically, the first and second excited states
correspond to the breakdown of the bound state at ${-(M-1)\eta}$, with
the first excited state in the antisymmetric subspace and the second
excited state in the symmetric subspace. Explicitly, we can write down
the spectral parameter configurations for the first two excited states
\bea
 v_{1+} &\ap& -M \eta +a_{1+} \eta ^{-3},~~   
 v_{1-} \ap -(M-1) \eta +a_{1-} \eta ^{-3},\no\\
 v_{j\pm} &\ap& -(M-j) \eta +a_{j\pm} \eta ^{-(2j-1)},
  ~~~~j=2,\cdots,M, \no
\eea
with
\bea
a_{1+} &=& -\frac {M+1}{2},~~~~   
a_{1-} = \frac {M(M+1)}{2}, \no\\
a_{2\pm} &=& \frac {-(M-1)^2 \pm (M-1) \sqrt {13 M^2 +10M +1}}{12}, \no\\
a_{3\pm} &=& \pm \frac {(M-1)(M-2) \sqrt {2M(M+1)}}{24}, \no\\
a_{j\pm} &=& \frac {M-j+1}{\sqrt {(j+1)j(j-1)(j-2)}} a_{j-1,\pm}, 
~~j=3,\cdots,M, \no
\eea
for the (antisymmetric) first excited state and
\bea
a_{1+} &=& -\frac {(M+1)(2M+1)}{2},~~~~ 
a_{1-} = -\frac {M(M+1)}{2}, \no\\
a_{2\pm} &=& \frac {-(M-1)^2 \pm i (M-1) \sqrt {11 M^2 +14M -1}}{12}, \no\\
a_{3\pm} &=& \pm i \frac {(M-1)(M-2) \sqrt {2M(M+1)}}{24}, \no\\
a_{j\pm} &=& \frac {M-j+1}{\sqrt {(j+1)j(j-1)(j-2)}} a_{j-1,\pm}, 
~~j=3,\cdots,M, \no
\eea
for the (symmetric) second excited state.
The breakdown of the bound state at ${-(M-j)\eta}$,  
$j=2,\cdots,M$ results in
the higher excited states.

Substituting these results into (\ref{becnrg})
leads us to the asymptotic ground state energy 
$$
E_0 \ap - 2\K \eta ^{-2}M(M+1),
$$
while for the first and second excited states we have
\bea
E_1 &\ap& \K \eta ^2  - \K \eta ^{-2} \frac {M^2 +M -2}{3},\no\\
E_2 &\ap& \K \eta ^2  + \K \eta ^{-2} \frac {5M^2 +5M +2}{3}.\no
\eea
In contrast to the Rabi regime, the Fock regime is not semiclassical, as 
the ratio of the gap $\Delta$ and $N$ is of finite order when
$N$ is large. 

We can perform a similar analysis for odd $N$. In this case, the gap
between the ground and the first excited states is proportional to
$\K \eta ^{-2}$ instead of $\K \eta ^2$. 
Furthermore, the ground
state root structure is different in the odd case since not all the
bosons can be bound in pairs. This indicates there
is a strong parity effect in the Fock regime, in contrast to the Rabi
regime. 

(ii) $\o \neq 0$. In this case the root structure is somewhat more 
complicated than for $\o=0$, so we will not present the details.
We remark however that our 
calculations show that up to order $\eta^{-2}$ the ground 
state energy eigenvalue takes the
same form as in the case $\o =0$. Actually, the leading contribution 
arising from
the $\o$ term appears only as $\o^2 \eta ^{-4}$.
This means that the results presented below are applicable for all
values of $\o$ (or equivalently $\Delta\mu$). 

Although it is difficult to define rigorously \cite{l01,phase}, 
the relative phase between Bose-Einstein condensates is useful
in understanding
interference experiments \cite{otfyk,cbfmmtsi,hmwc98}. 
Recall that in Josephson's original proposal \cite{jo} 
for Cooper pair tunneling 
through an insulating barrier between 
macroscopic superconductors, the current is a manifestation of the 
relative phase between the wavefunctions of the superconductors. 
By definition, the relative 
phase $\Phi$ is
conjugate to the relative number of atoms in the two condensates
$n\equiv N_1-N_2$. Using the Hellmann-Feynman theorem, we find that
$$<\Delta n^2> = 8 \frac {\partial E_0}{\partial K} -4  \left( 
 \frac {\partial E_0}{\partial \Delta \mu} \right)^2.$$ 
For the ground state 
in the limit of strong  tunneling (i.e., Rabi regime),
$<\Delta n^2> \ap N-({\Delta \mu N}/{{\cal E}_J})^2$. 
In the case of weak  tunneling (i.e., Fock regime),
$<\Delta n^2> \ap 2 N(N+2) ( {{\cal E}_J}/{K})^2$. 
The degree of coherence between the two Bose-Einstein 
condensates can be discussed in terms of
\cite{l01} 
$$\a \equiv \frac{1}{2N}
<a_1^\dagger a_2 + a_2^\dagger a_1> = - \frac{1}{N} 
\frac{\partial E_0}{\partial
{\cal E}_J }.$$ 
In the strong coupling limit, $\a \ap 1- N^{-1} (\Delta
\mu)^2 / (8 {\cal E}_J)^2$, indicating very close to
full coherence in the ground state. 
In the opposite limit, we have $\a \ap 2(N+2) {\cal
E}_J / K<<1$, indicating the absence of coherence.
The above results give the first order corrections to the results 
presented in 
\cite{otfyk,ps01} for the number
fluctuations and the coherence factor at zero temperature.

\section{A model for atomic-molecular Bose-Einstein condensation}


\def\aa{\alpha} 
\def\bb{\beta}
\def\a{\hat a}
\def\b{\hat b}
\def\d{\dagger}
\def\de{\delta} 
\def\e{\epsilon}
\def\g{\gamma}
\def\K{\kappa}
\def\ap{\approx}
\def\l{\lambda}
\def\o{\omega}
\def\t{\tilde{\tau}}
\def\s{S}
\def\D{\Delta}
\def\L{\Lambda}
\def\T{{\cal T}}
\def\TT{{\tilde{\cal T}}}
\def\E{{\cal E}} 

\def\beq{\begin{equation}}
\def\eeq{\end{equation}}
\def\bea{\begin{eqnarray}}
\def\eea{\end{eqnarray}}
\def\ba{\begin{array}}
\def\ea{\end{array}}
\def\no{\nonumber}
\def\le{\langle}
\def\re{\rangle}
\def\lt{\left}
\def\rt{\right}
\def\o{\omega}
\def\d{\dagger}
\def\nn{\nonumber}
\def\j{{ {\cal J}}}
\def\n{{\hat n}}
\def\N{{\hat N}}
\def\T{{\cal T}}
\def\TT{{\tilde {\cal T}}}

After the experimental realisation of Bose-Einstein condensation
in
dilute alkali gases, many physicists started to 
consider the possibility of producing a molecular Bose-Einstein
condensate from
photoassociation and/or the Feshbach resonance
of an atomic Bose-Einstein condensate of a weakly interacting dilute
alkali gas
\cite{wynar,inouye}. This novel area
has attracted  considerable attention from both experimental and
theoretical
physicists, 
and in particular it has recently been reported that a Bose-Einstein 
condensate of rubidium has been achieved comprised of a coherent superposition 
of atomic and molecular states \cite{zoller,donley}.
As stressed in 
\cite{vardi}, even in the ideal two-mode limit, mean field theory
fails to provide long-term predictions due to strong interparticle
entanglement
near the dynamically unstable molecular mode.
The numerical results have shown that the large-amplitude atom-molecular
coherent oscillations are damped by the rapid growth
of fluctuations near the unstable point, which
contradicts
the mean field theory predictions. In order to clarify the 
controversies raised by these investigations, one can appeal to the 
exact solution of the two mode model, the derivation of which we will 
now present. 

The two mode Hamiltonian takes the form
\beq
H= \frac {\omega}{2} {a}^\dagger {a} +\frac {\Omega}{2}
 ({a}^\dagger {a}^\dagger {b}
  +{b}^\dagger {a}{a}),
  \label{abham}
  \eeq
  where ${a}^\dagger$ and ${b}^\dagger$ denote the creation
  operators
  for atomic and molecular modes respectively.
  Note that the total atom number operator
$\N= N_a+2N_b$ where 
$N_a={a}^\dagger {a},\,N_b=  {b}^\dagger {b}$ provides
  a good quantum number since $[H,\,{\hat N}]=0.$

In order to derive this Hamiltonian through the quantum inverse 
scattering method, we take the following $L$-operator 
$$L(u)=GL^b(u-\delta-\eta^{-1})L^K(u)$$ 
with the matrix $G$ given by 
$$G=\pmatrix{-\eta^{-1} & 0 \cr 0 & \eta^{-1}}. $$ 
This gives us the explicit realisation of the Yang-Baxter algebra 
\bea 
A(u)&=&-\eta^{-1}(u+\eta K^z)(u-\delta-\eta^{-1}+\eta N_b)+bK^+    \nn \\
B(u)&=& 
- K^-(u-\delta-\eta^{-1}+N_b)
-\eta^{-1}b(u-\eta K^z) \nn \\
C(u)&=& \eta^{-1}b^\d(u+\eta K^z)-\eta^{-1}K^+    \nn \\
D(u)&=& b^\d K^- +\eta^{-2}(u-\eta K^z)   \nn \eea 
and 
\beq t(0)=\delta K^z+b^\d K^-+bK^+ -\eta  K^z N_b. \label{t0} \eeq   

Let $\left|0\right>$ denote the Fock vacuum state and let $\left|k\right>$ 
denote a lowest weight state of the $su(1,1)$ algebra with weight $k$; i.e.,
$K^z\left|k\right>=k\left|k\right>$. On the product state 
$\left|\Psi\right>=\left|0\right>
\left|k\right>$ it is clear that $B(u)\left|\Psi\right>=0$ and  
\bea a(u)&=&-\eta^{-1}(u+\eta k)(u-\delta-\eta^{-1}) \nn \\
d(u)&=&\eta^{-2}(u-\eta k).  \nn \eea 
We can immediately conclude that the eigenvalues of (\ref{t0}) are given by 
\beq \Lambda(0)=k(\delta+\eta^{-1})\prod_{i=1}^M\frac{v_i-\eta}{v_i}
-k\eta^{-1}\prod_{i=1}^M\frac{v_i+\eta}{v_i}    
\eeq 
subject to the Bethe ansatz equations 
\beq\frac{(v_i+\eta k)(1-\eta v_i+\eta \delta)}{(v_i-\eta k)}
=\prod_{j\neq i}^M\frac{v_i-v_j-\eta}{v_i-v_j+\eta}.  
\label{qambae}\eeq 

Realising the $su(1,1)$ algebra in terms of canonical boson operators through 
$$K^+=\frac{(a^\d)^2}{2},~~~K^-=\frac{a^2}{2},~~~K^z=\frac{2N_a+1}{4}$$ 
we then find that the Hamiltonian (\ref{abham}) is related to (\ref{t0}) 
through 
$$H=\lim_{\eta\rightarrow 0}\Omega(t(0)-\delta/4)$$ 
with $\omega=\Omega\delta$. Note that in this case the possible lowest weight 
states for the $su(1,1)$ algebra are  
$$\left|k=1/4\right>\equiv \left|0\right>,~~~\left|k=3/4\right>\equiv a^\d
\left|0\right>. $$  
Moreover, we have $N=2M+2k-1/2.$ 

It is worth mentioning at this point that another realisation of the 
$su(1,1)$ algebra is given in terms of two sets of boson operators by 
$$K^+=a^\d c^\d,~~~K^-=ac,~~~K^z=\frac{N_a+N_c+1}{2} $$ 
with $J=N_a-N_c$ a central element commuting with the $su(1,1)$ algebra 
in this representation.  
Due to the symmetry $a^\d\leftrightarrow c^\d$ we may assume $J\geq 0$.
For this case we define the Hamiltonian 
\bea H&=&  \lim_{\eta\rightarrow 0}
\Omega(t(0)-\delta/2)+\beta J\nn \\
&=& \alpha N_a+ \gamma N_c +\Omega(a^\d c^\d b+ b^\d a c) 
\label{abcham} 
\eea 
with $\alpha=\delta\Omega/2+\beta$ and $\gamma=\delta\Omega/2-\beta$. 
This model has a natural interpretation for atomic-molecular
Bose-Einstein condensation for two distinct atomic species which can bond 
to form a di-atomic molecule. In this case the possible lowest weight
states for the $su(1,1)$ algebra are 
$$\left|k=(m+1)/2\right>\equiv \left(a^\d\right)^m\left|0\right> $$ 
and $J=2k-1$. 
A detailed analysis of this model through the exact solution will be given 
at a later date.         

For the exact solution of the Hamiltonian (\ref{abham}) it is 
necessary  to take the 
{\it quasi-classical limit} $\eta\rightarrow 0$ in the Bethe ansatz
equations (\ref{qambae}).  
The resulting Bethe ansatz 
equations take the form 
\beq
\delta - v_i + \frac {2k}{v_i} = 2 \sum ^M_{j \neq i} \frac {1}
{v_j -v_i}. \label {abbae}
\eeq
Also, in this limit    
the corresponding energy eigenvalue is 
\bea {E}&=&\omega( M+k-1/4)-\Omega\sum_{i=1}^M v_i \nn \\
&=&\omega(k-1/4)-2k\Omega\sum_{i=1}^M\frac{1}{v_i}.  \label{amnrg} \eea 
The equivalence of the two energy expressions can be deduced from
(\ref{abbae}). 
The eigenstates too are
obtained by this procedure. Consider the following
class of states 
\beq \left|v_1,...,v_M\right>=\prod_{i=1}^Mc(v_i)\left| 
\Psi\right> \label{estates} \eeq  
where $c(v)=(vb^\d-a^\d a^\d/2)$, 
$\left|\Psi\right>=\left|0\right>$ for $k=1/4$ and
$\left|\Psi\right>=a^\d\left|0\right>$ for $k=3/4$. 
In the case when the set of parameters $\{v_i\}$ satisfy the Bethe ansatz 
equations (\ref{abbae}),  then
(\ref{estates}) 
are precisely the eigenstates of the Hamiltonian. 

\subsection{Asymptotic analysis of the solution} 
In the limit of large $|\de|$ we can perform
an asymptotic analysis of the Bethe ansatz equations to determine the
asymptotic form of the energy spectrum.  
We choose the following ansatz for the Bethe roots
\bea v_i&\ap& \delta^{-1}\mu_i  ~~~~~~~~~~~~~~~~~~~~~~~~i\leq m, \nn \\
v_i&\ap&\delta +\e_i+\delta^{-1}\mu_i ~~~~~~~~~~~~~~~i>m. \nn \eea
For $i>m$ we obtain  from the zero order terms in the Bethe ansatz equations
$$\e_i=2\sum_{\stackrel{j=m+1}{j\neq i}}^M\frac{1}{\e_i-\e_j} $$
which implies
$$\sum_{i=m+1}^M \e_i=0.$$  
{}From the terms in $\delta^{-1}$ we
find
$$\mu_i=2(k+m)+2\sum_{\stackrel{j=m+1}{j\neq i}}^M
\frac{\mu_j-\mu_i}{(\e_j-\e_i)^2}$$
and thus
$$\sum_{i=m+1}^M \mu_i=2(k+m)(M-m).$$ 

Next we look at the Bethe ansatz equations
 for $i\leq m$. The terms in $\delta$ give
$$1+\frac{2k}{\mu_i}=2\sum_{j\neq i}^m\frac{1}{\mu_j-\mu_i}$$
which implies
$$\sum_{i=1}^m \mu_{i}=-2km-m(m-1).$$  
This gives the energy levels
\bea
{E_m}&\ap& \omega( M+(k-1/4))-\omega(M-m)
-\Omega\sum_{i=m+1}^M\e_i-\frac{\Omega^2}{\omega}\sum_{i=1}^M\mu_i \nn \\
&=&\omega(m+k-1/4)   
+
\frac{\Omega^2}{\omega}(3m^2-m+4km-2kM-2mM).   \nn
\eea
The level spacings are
\bea \Delta_m&=&E_m-E_{m-1} \nn \\
&\ap& \omega-\frac{2\Omega^2}{\omega}(M+2-3m-2k) \nn
\eea
from which we conclude that in this limit the model is semi-classical.

Let $\E$ denote the ground state energy ($\E=E_0$ for $\Omega\delta>>0$, 
$\E=E_M$ for $\Omega\delta<<0$) and $\Delta$ the
gap to the first excited state. Employing the Hellmann-Ferynman theorem
we can determine the asymptotic form of the following zero temperature
correlations 
$$\left<N_a\right>=2\frac{\partial\E}{\partial \omega}, ~~~~
\theta=-2\frac{\partial \E}{\partial \Omega}  $$   
where $\theta=-\left<{a}^\dagger {a}^\dagger {b}
  +{b}^\dagger {a}{a}\right>$ 
  is the coherence correlator. 
For large $N$ we introduce the rescaled variables 
\bea 
\delta^*=\frac{\delta}{N^{1/2}},
&~~~~~&\Delta^*=\frac{\Delta}{\Omega N^{1/2}},~~~~~    
\left<N_a\right>^*=\frac{\left<N_a\right>}{N} , 
~~~~~\theta^*=\frac{\theta}{N^{3/2}}. 
\label{scale} \eea 
We then have for  $\delta^*>> 0$   
$$ 
\Delta^*\ap\delta^*-\frac{1}{\delta^*},~~~     
\left<N_a\right>^*\ap 0, ~~~    
\theta^*\ap 0   $$       
while for $\delta^*<<0$ 
$$ 
\Delta^*\ap -\delta^*-\frac{2}{\delta^*}, ~~      
\left<N_a\right>^*\ap 1-\frac{1}{2(\delta^*)^2},~~     
\theta^*\ap -\frac{1}{\delta^*}       .  $$ 
The above shows that the model has scale invariance in the asymptotic
limit. The scaling properties actually hold for a wide range of 
values of the 
scaled detuning parameter $\delta^*$, which is established through numerical 
analysis \cite{zlm}. 

\subsection{Computing the energy spectrum} 
For this model 
there is a convenient method to determine the energy spectrum without
solving the Bethe ansatz equations (cf. \cite{rybin}). 
This is achieved by introducing the polynomial function
whose zeros are the roots of the Bethe ansatz equations; viz.
$$G(u)=\prod_{i=1}^M(1-u/v_i). $$
It can be shown from the Bethe ansatz equations that $G$
satisfies the differential equation
\beq uG''-(u^2-\delta u-2k)G'+(Mu-E/\Omega+\delta(k-1/4))G=0 \label{de} \eeq 
subject to  the initial conditions
$$G(0)=1,~~~~G'(0)=\frac{E-\omega(k-1/4)}{2k\Omega}. $$
In order to show this we set 
$$F(u)=uG''-(u^2-\delta u-2k)G'. $$ 
As a result of the Bethe ansatz equations (\ref{abbae}) it is deduced that 
$F(v_i)=0$. Given that $F(u)$ is a polynomial of degree $(M+1)$, we then 
conclude that $F(u)=(\alpha u+\beta)G(u)$ for some constants $\alpha, 
\, \beta$, which are determined by the asymptotic limits $u\rightarrow 0$ 
and $u\rightarrow\infty$. Eq. (\ref{de}) then follows.

By setting $G(u)=\sum_n g_n u^n $ the recurrence relation
\beq g_{n+1}=\frac{E-\omega (n+k-1/4)}{\Omega (n+1)(n+2k)} g_n
+\frac{n-M-1}{(n+1)(n+2k)}
g_{n-1} \label{rec} \eeq
is readily obtained. It is clear from this relation that $g_n$ is a
polynomial in $E$ of
degree $n$. We also know that $G$ is a polynomial function of degree
$M$ and so we must have $g_{M+1}=0$. The $(M+1)$ roots of $g_{M+1}$
are precisely the energy levels $E_m$.
Moreover, the eigenstates (\ref{estates}) are expressible as (up to
overall normalisation)
$$\left|v_1,...,v_M\right>=\sum_{n=1}^Mg_n(b^\d)^{(M-n)}
\left(\frac{a^\d a^\d}{2}\right)^n
\left|\Psi\right>. $$

The recurrence relation (\ref{rec}) can be solved as follows 
(cf. \cite{rybin}).
Setting
$$
g_{n+1} = g_0 \prod ^n_{j=0} x_j y_j,
$$
with
$$
x_j = \frac{E-\omega (j +k-1/4)}{\Omega (j+1)(j+2k)},
$$
and substituting into the recurrence relation (\ref{rec}), we have
$$
x_j x_{j-1} y_{j-1} (y_{j}-1) = \frac {j-M-1}{(j+1)(j+2k)}.
$$
This yields
$
y_j = 1+ c_{j-1}/y_{j-1}
$
with
$$
c_j = \frac {\Omega^2 (j+1)(j+2k)(j-M)}
{(E-\omega (j+k+3/4))(E-\omega (j+k-1/4))}, 
$$
which means $y_j$ can be expressed as  a continued fraction. 
The requirement that $G$ is a polynomial function of order $M$ decrees
$y_M=0$, in turn implying 
$$ y_{M-1} = \frac {\Omega^2 M(M+2k-1)}
{(E-\omega (M+k-1/4))(E-\omega (n+k-5/4))},
$$
which is an algebraic equation that determines the allowed energy
levels $E_m$. 
The above procedure can easily be employed to determine the energy
spectrum numerically, without resorting to solving the Bethe ansatz
equations. Explicit results can be found in \cite{zlm}.

\section{The BCS Hamiltonian}

\def\beq{\begin{equation}}
\def\eeq{\end{equation}}
\def\bea{\begin{eqnarray}}
\def\eea{\end{eqnarray}}
\def\ve{\epsilon}
\def\si{\sigma}
\def\th{\theta}
\def\d{\delta}
\def\ga{\gamma}
\def\l{\left}
\def\r{\right}
\def\a{\alpha}
\def\b{\beta}
\def\g{\gamma}
\def\La{\Lambda}
\def\w{\overline{w}}
\def\u{\overline{u}}
\def\o{\overline}
\def\rr{\mathcal{R}}
\def\T{\mathcal{T}}
\def\N{\overline{N}}
\def\Q{\overline{Q}}
\def\L{\mathcal{L}}
\def\m{\overline{m}}
\def\n{\overline{n}}
\def\p{\overline{p}}
\def\l{\overline{l}}
\def\d{\dagger}


\def\a{\alpha}
\def\b{\beta}
\def\d{\dagger}
\def\e{\epsilon}
\def\g{\gamma}
\def\k{\kappa}
\def\l{\lambda}
\def\o{\omega}
\def\t{\tilde{\tau}}
\def\s{S}
\def\D{\Delta}
\def\T{{\cal T}}
\def\TT{{\tilde{\cal T}}}

\def\beq{\begin{equation}}
\def\eeq{\end{equation}}
\def\bea{\begin{eqnarray}}
\def\eea{\end{eqnarray}}
\def\ba{\begin{array}}
\def\ea{\end{array}}
\def\no{\nonumber}
\def\le{\langle}
\def\re{\rangle}
\def\lt{\left}
\def\rt{\right}
\def\oR{R^*} \def\upa{\uparrow}
\def\R{\overline{R}} \def\doa{\downarrow}
\def\oL{\overline{\Lambda}}
\def\nn{\nonumber} \def\dag{\dagger}
\def\e{\epsilon}
\def\si{\sigma}
\def\th{\theta}
\def\de{\delta}
\def\ga{\gamma}
\def\l{\left}
\def\r{\right}
\def\a{\alpha}
\def\b{\beta}
\def\g{\gamma}
\def\La{\Lambda}
\def\w{\overline{w}}
\def\u{\overline{u}}
\def\o{\overline}
\def\rr{\mathcal{R}}
\def\T{\mathcal{T}}
\def\N{\overline{N}}
\def\Q{\overline{Q}}
\def\L{\mathcal{L}}
\def\k{\overline{k}}
\def\l{\overline{l}}
\def\d{\dagger}

The experimental work of Ralph, Black and Tinkham \cite{rbt} 
on the discrete 
energy spectrum in small metallic aluminium grains generated 
interest in understanding the nature of superconducting correlations at the 
nanoscale level. Their results indicate significant parity effects due to the 
number of electrons in the system. For grains with an odd number of electrons,
the gap in the energy spectrum reduces with the size of the system, in contrast 
to the case of a grain with an even number of electrons, where a gap 
larger than the single electron energy levels persists. In the latter case 
the gap can be closed by a strong applied magnetic field. 
The conclusion drawn from these results is that pairing interactions
are prominent in these nanoscale systems.
For a grain with an odd number of electrons there will always 
be at least one unpaired electron, so it is not necessary to break a Cooper
pair in order to create an excited state. For a grain with an even number 
of electrons, all excited states have a least one broken Cooper pair,
resulting in a gap in the spectrum.
In the presence of a 
strongly applied magnetic field, it is energetically more favourable 
for a grain with an even number of electrons 
to have broken pairs, and hence in this case 
there are excitations which show no gap in the 
spectrum. 
    
The physical properties of a small metallic grain are described by the
reduced BCS Hamiltonian \cite{vdr}   
\beq H=\sum_{j=1}^{\L}\e_jn_j
-g\sum_{j,k}^{\L}c_{k+}^{\d}c_{k-}^{\d}c_{j-}c_{j+}. \label{bcs}
\eeq 
Above, $j=1,...,{\L}$ labels a shell of doubly degenerate single particle
energy levels with energies $\e_j$ and $n_j$ is the 
fermion number operator for
level $j$. The operators $c_{j\pm},\,c^{\d}_{j\pm}$ are the annihilation
and creation operators for the fermions at level $j$. The labels $\pm$ refer 
to time reversed states.  

One of the features of the Hamiltonian (\ref{bcs}) 
is the {\it blocking
effect}. For any unpaired electron at level $j$ the action of 
the pairing interaction is zero since only paired electrons are
scattered. This means that the Hilbert space can be decoupled into 
a product of paired and unpaired electron states in which the
action of the Hamiltonian on the subspace for the unpaired electrons is 
automatically diagonal in the natural basis. 
In view of the blocking effect, it is convenient to introduce 
hard-core boson operators 
$b_j=c_{j-}c_{j+},\, b^{\d}_{j}=c^{\d}_{j+}c^{\d}_{j-}$ which satisfy
the relations 
$$(b^{\d}_j)^2=0, ~~~~[b_j,\,b_k^{\d}]=\delta_{jk}(1-2b^{\d}_jb_j) 
~~~~[b_j,\,b_k]=[b^{\d}_j,\,b^{\d}_k]=0 $$
on the subspace excluding single particle states. 
In this setting the hard-core boson operators realise the
$su(2)$ algebra in the pseudo-spin reprepresentation, 
which will be utilised below.

The original approach of Bardeen, Cooper and Schrieffer \cite{bcs} 
to describe the 
phenomenon of superconductivity was to employ
a mean field theory using a variational wavefunction for the ground state 
which has an undetermined number of electrons. The expectation value for the 
number operator is then fixed by means of a chemical potential term $\mu$. 
One of the 
predictions of the BCS theory is that the number of Cooper pairs in the ground 
state of the system is given by the ratio $\Delta/d$ where $\Delta$ is the BCS 
``bulk gap'' and $d$ is the mean level spacing for the single electron 
eigenstates. For nanoscale systems, this ratio is of the order of unity, in 
seeming contradiction with the experimental results discussed above. 
The explanation for this is that the mean-field approach is inappropriate
for nanoscale systems due to large superconducting fluctuations.

As an alternative to the BCS mean field approach, one can appeal to the 
exact solution of the Hamiltonian (\ref{bcs}) derived by Richardson 
\cite{richardson}
and developed by Richardson   and Sherman \cite{rs}.
It has also been shown by Cambiaggio, Rivas and Saraceno \cite{crs} that 
(\ref{bcs})
is integrable in the sense that there exists a set of mutually commutative 
operators which commute with the Hamiltonian. These features have recently been 
shown to be a consequence of the fact that the model can be derived in the 
context of the quantum inverse scattering method using the $L$-operator 
(\ref{ls}) with a $c$-number $L$-operator 
\cite{zlmg,vp}, which we will now explicate.

\subsection{A universally integrable system}

In this case we use 
a $c$-number realisation $G$ of the $L$-operator as well as 
(\ref{ls}) 
to  construct the transfer matrix 
\beq t(u)={\rm tr}_0\left(G_0L_{0{\L}}(u-\e_{\L})...L_{01}(u-\e_1)\right) 
\label{bcstm} \eeq  
which is an element of the $\L$-fold tensor algebra of $su(2)$.
Above, 
${\rm tr}_0$ denotes the trace
taken over the auxiliary space labelled 0 
and $G=\exp(-\alpha\eta {\sigma})$ with
$\sigma={\rm diag}(1,\,-1)$. 
Defining 
$$T_j=\lim_{u\rightarrow \e_j}\frac{u-\e_j}{\eta^2} t(u) $$ 
for $j=1,2,...,{\L}$,
we may write in the quasi-classical limit  
$T_j=\tau_j+o(\eta) $,   
and it follows from the commutivity of the transfer matrices that 
$ [\tau_j,\,\tau_k]=0, ~ \forall\, j,\,k. $ 
Explicitly, these operators read 
\beq 
\tau_j=2\alpha S^z_j+\sum_{k\neq j}^{\L}\frac{\theta_{jk}}{\e_j-\e_k}
\label{cons} \eeq 
with 
$\theta=S^+\otimes S^-+S^-\otimes S^++2S^z\otimes S^z.$ 

We define a Hamiltonian through
\bea  
H&=&-\frac{1}{\a}\sum_{j=1}^{\L}\e_j\tau_j
+\frac{1}{4\a^3}\sum_{j,k=1}^{\L}\tau_j\tau_k 
+\frac{1}{2\a^2}\sum_{j=1}^{\L} \tau_j- \frac{1}{2\a}\sum_{j=1}^{\L}C_j 
\\ 
&=& -\sum_{j=1}^{\L}2\e_jS_j^z-\frac{1}{\a}\sum_{j,k=1}^{\L}S_j^-S_k^+
\label{ham1} \eea       
where 
$$C=S^+S^-+S^-S^++2(S^z)^2$$ 
is the Casimir invariant for the $su(2)$ algebra. 
The Hamiltonian is universally integrable since it is clear that 
$[H,\,\tau_j]=0,~~\forall j $ irrespective of the realisations of the 
$su(2)$ algebra in the tensor algebra.  

In order to reproduce the Hamiltonian (\ref{bcs}) we realise 
the $su(2)$ generators through the hard-core boson  
(spin 1/2) representation; viz
\beq S_j^+=b_j, ~~~
S_j^-=b_j^{\dagger}, ~~~ 
S^z_j=\frac{1}{2}\left(I-n_{j}\right) \label{psr}. \eeq  
In this instance one obtains (\ref{bcs})
(with the  constant term $-\sum_j^\L\e_j$) where $g=1/\a$ 
as shown by Zhou et al. \cite{zlmg}
and von Delft and Poghossian \cite{vp}. 

For each index $k$ in the tensor algebra in which the transfer matrix acts, 
and accordingly in (\ref{ham1}),
suppose that we represent the $su(2)$ 
algebra through the irreducible representation with spin $s_k$. 
Thus $\{S^+_k,\,S^-_k,\,S^z_k\}$ act on a $(2s_k+1)$-dimensional space. 
In employing the method of the algebraic Bethe ansatz 
discussed earlier we find that 
\bea a(u)&=&\exp(-\a\eta)\prod_{k=1}^\L\frac{u-\e_k-\eta s_k}
{u-\e_k} \nn \\
d(u)&=&\exp(\a\eta)\prod_{k=1}^\L\frac{u-\e_k+\eta s_k}{u-\e_k} \nn \eea 
which gives the eigenvalues of the transfer matrix  
(\ref{bcstm}) as 
\bea
\Lambda(u)&=&\exp(\alpha\eta)\prod_{k=1}^{\L}\frac{u-\e_k+\eta s_k}{u-\e_k}
\prod_{j=1}^M\frac{u-v_j-\eta}{u-v_j} \nn \\
&&~
+\exp(-\alpha\eta)\prod_{k=1}^{\L}\frac{u-\e_k-\eta s_k}{u-\e_k}
\prod_{j=1}^M\frac{u-v_j+\eta}{u-v_j}. \nn  
 \eea
The corresponding Bethe ansatz equations read 
$$\exp(2\alpha\eta)\prod_{k=1}^{\L}\frac{v_l-\e_k+\eta s_k}
{v_l-\e_k-\eta s_k}
=-\prod_{j=1}^M\frac{v_l-v_j+\eta}{v_l-v_j-\eta} . $$

The eigenvalues of the conserved operators (\ref{cons}) are obtained
through the appropriate 
terms in the expansion of the transfer matrix eigenvalues  
in the parameter $\eta$. This yields the following result 
for the eigenvalues $\lambda_j$ of $\tau_j$ 
\beq \lambda_j= \left(2\a +\sum_{k\neq j}^{\L}\frac{2s_k}{\e_j-\e_k} 
-\sum_{i=1}^M 
\frac{2}{\e_j-v_i}\right)s_j \label{eig} \eeq  
such  that the parameters $v_j$ now satisfy the Bethe ansatz equations 
\beq 
2\a+ \sum_{k=1}^{\L}\frac{2s_k}{v_j-\e_k}
=\sum_{i\neq j}^M \frac{2}{v_j-v_i}.
\label{bcsbae} \eeq  
Through (\ref{eig}) we can now determine the energy eigenvalues of 
(\ref{ham1}). It is useful to note the following identities 
\bea 
2\a \sum_{j=1}^{M}v_j+2\sum_{j=1}^M\sum_{k=1}^{\L}\frac{v_js_k}{v_j-\e_k}
&=&M(M-1) \nn \\
\a M+\sum_{j=1}^M\sum_{k=1}^{\L}\frac{s_k}{v_j-\e_k}&=&0 \nn \\
\sum_{j=1}^M\sum_{k=1}^{\L}\frac{v_js_k}{v_j-\e_k} -
\sum_{j=1}^M\sum_{k=1}^{\L}\frac{s_k\e_k}{v_j-\e_k}&=&M\sum_{k=1}^{\L}s_k. 
\nn\eea 
Employing the above it is deduced that 
\bea 
\sum_{j=1}^{\L}\lambda_j &=&2\a\sum_{j=1}^{\L}s_j-2\a M \nn \\
\sum_{j=1}^{\L}\e_j\lambda_j&=&2\a \sum_{j=1}^{\L}\e_j s_j
+\sum_{j=1}^{\L}\sum_{k\neq j}^{\L}  
s_js_k-2M\sum_{k=1}^{\L}s_k-2\a\sum_{j=1}^M v_j +M(M-1) \nn \eea 
which, combined with the eigenvalues $2s_j(s_j+1)$ for the Casimir invariants
$C_j$, yields the energy eigenvalues 
\beq E=2\sum_{j=1}^{M} v_j. \label{bcsnrg} \eeq  
{}From the above expression we see that the quasi-particle excitation
energies are given by twice the Bethe ansatz roots $\{v_j\}$ 
of (\ref{bcsbae}). 
In order to specialise this result to the case of the BCS 
Hamiltonian (\ref{bcs}) it is a matter of setting $s_k=1/2,~\forall\,k$. 
Finally, let us remark that in the quasi-classical limit the eigenstates 
assume the form 
$$\left|\Psi\right>=\prod_{i=1}^M\sum_{j=1}^\L\frac{b_j^\d}{v_i-\e_j}
\left|0\right>. $$

The construction given above can also be applied on a more general level. 
Taking 
higher spin representations of the $su(2)$ algebra produces models of 
BCS systems which are coupled by Josephson tunneling, as described in 
\cite{lzmg,lh}. One can also employ higher rank Lie algebras, such as $so(5)$
\cite{lzgm} and $su(4)$ \cite{gflz} which produce coupled BCS systems which 
model pairing interactions in nuclear systems. For the general case of an
arbitrary Lie algebra we refer to \cite{afs}. Finally, let us mention that if 
one reproduces the above construction with the $su(1,1)$ $L$-operator (\ref{lk})
in place of the $su(2)$ $L$-operator (\ref{ls}) the pairing model for 
bosonic systems introduced by Dukelsky and Schuck \cite{ds} is obtained.   
\def\oR{R^*} \def\upa{\uparrow}
\def\R{\overline{R}} \def\doa{\downarrow}
\def\oL{\overline{\Lambda}}
\def\nn{\nonumber} \def\dag{\dagger}
\def\beq{\begin{equation}}
\def\eeq{\end{equation}}
\def\bea{\begin{eqnarray}}
\def\eea{\end{eqnarray}}
\def\ve{\epsilon}
\def\si{\sigma}
\def\th{\theta}
\def\d{\delta}
\def\ga{\gamma}
\def\l{\left}
\def\r{\right}
\def\a{\alpha}
\def\b{\beta}
\def\g{\gamma}
\def\La{\Lambda}
\def\w{\overline{w}}
\def\u{\overline{u}}
\def\o{\overline}
\def\rr{\mathcal{R}}
\def\T{\mathcal{T}}
\def\N{\overline{N}}
\def\Q{\overline{Q}}
\def\L{\mathcal{L}}
\def\m{\overline{m}}
\def\n{\overline{n}}
\def\p{\overline{p}}
\def\l{\overline{l}}
\def\d{\dagger}
\subsection{Asymptotic analysis of the solution}

In the limit $g\rightarrow 0$ we can easily determine the ground state
energy of (\ref{bcs}); it is given by filling the Fermi sea. 
Below we will assume that the
number of fermions is even. 
Thus for small $g>0$ it is appropriate to consider the
asymptotic solution 
$$v_i\ap \e_i+g\delta_i+g^2\mu_i, ~~~i=1,...,M. $$ 
Substituting this into (\ref{bcsbae}) 
and equating the different orders in $g$
yields 
$$v_i\ap \e_i-\frac{g}{2}+\frac{g^2}{4}\left(\sum_{k=m+1}^\L\frac{1}{\e_j-\e_k}
-\sum_{i\neq j}^M \frac{1}{\e_j-\e_i}\right) $$ 
which immediately gives us the
asymptotic ground state energy 
$$E_0\ap 2\sum_{j=1}^M\e_j-gM+\frac{g^2}{2}\sum_{j=1}^M\sum_{k=M+1}^\L
\frac{1}{\e_j-\e_k} . $$   

Next we look at the first excited state. In the $g=0$ case this
corresponds to breaking the Cooper pair at level $\e_M$ and putting single
unpaired electrons in the levels $\e_M$ and $\e_{M+1}$. Now these two
levels become blocked. To solve the equations (\ref{bcsbae})
 for this  excited state is the
same as for the ground state except that there are now $(M-1)$ Cooper pairs
and we have to exclude the blocked levels. We can therefore write down
the energy 
$$E_1\ap \e_M+\e_{M+1}+ 2\sum_{j=1}^{M-1}\e_j -g(M-1)
+\frac{g^2}{2}\sum_{j=1}^{M-1}\sum_{k=M+2}^\L\frac{1}{\e_j-\e_k}. $$ 
The gap is found to be 
$$\Delta\ap \e_{M+1}-\e_M +{g}+\frac{g^2}{2}\left(\sum_{j=1}^{M-1}
\frac{1}{\e_{M+1}-\e_j}+\sum_{k=M+1}^\L\frac{1}{\e_k-\e_M}\right). $$ 

As in previous examples, we can calculate some asymptotic correlation
functions for zero temperature by using the Hellmann-Feynman theorem. 
In particular, 
$$\left<n_i\right>=\frac{\partial E_0}{\partial \e_i} $$ 
which for $i\leq M$ gives 
$$\left<n_i\right>\ap2-\frac{g^2}{2}\sum_{k=M+1}^\L\frac{1}{(\e_i-\e_k)^2}$$ 
while for $i>M$ we get 
$$\left<n_i\right>\ap\frac{g^2}{2}\sum_{j=1}^M\frac{1}{(\e_j-\e_i)^2}. $$ 
We can also determine the asymptotic form of 
the Penrose-Onsager-Yang Off-Diagonal Long-Range Order 
parameter \cite{po,y} to be  
\bea \frac{1}{\L}\sum_{i,j=1}^\L \left<b_i^\d b_j\right> 
&=&-\frac{1}{\L}\frac{\partial E_0}
{\partial g} \nn \\
&\ap&\frac{M}{\L}
-\frac{g}{\L}\sum_{j=1}^M\sum_{k=M+1}^\L
\frac{1}{\e_j-\e_k}. \nn \eea  

\begin{flushleft}
\bf{Acknowledgements}
\end{flushleft}
We are deeply indebted to 
 Ross McKenzie, Mark Gould, Xi-Wen Guan and Katrina Hibberd
for their collaborations on these topics. Financial support from the
Australian Research Council and Funda\c{c}\~{a}o de
Amparo \`a Pesquisa do Estado do Rio Grande do Sul is gratefully accepted.


\end{document}